\documentclass[]{aastex631}
\usepackage{float}
\usepackage{cancel}
\usepackage{soul}
\usepackage{graphicx}
\usepackage{textcomp}
\usepackage{comment}
\usepackage{ulem}
\usepackage{lipsum}
\usepackage{amsmath}
\usepackage{array}
\usepackage{booktabs}
\usepackage{color}
\usepackage{tabularx}

\begin{document}

\title{The discovery of three pulsars in the globular cluster M15 with the FAST}

\author{Yuxiao Wu}
\affiliation{School of Science, Chongqing University of Posts and Telecommunications \\
Chongqing, 40000, China}

\author{Zhichen Pan}
\affiliation{National Astronomical Observatories, Chinese Academy of Sciences, 20A Datun Road, Chaoyang District\\
Beijing, 100101, China}
\affiliation{CAS Key Laboratory of FAST, National Astronomical Observatories, Chinese Academy of Sciences\\
Beijing 100101, China}
\affiliation{College of Astronomy and Space Sciences, University of Chinese Academy of Sciences\\
Beijing 100049, China}

\author{Lei Qian}
\affiliation{National Astronomical Observatories, Chinese Academy of Sciences, 20A Datun Road, Chaoyang District\\
Beijing, 100101, China}
\affiliation{CAS Key Laboratory of FAST, National Astronomical Observatories, Chinese Academy of Sciences\\
Beijing 100101, China}
\affiliation{College of Astronomy and Space Sciences, University of Chinese Academy of Sciences\\
Beijing 100049, China}

\author[0000-0001-5799-9714]{Scott M. Ransom}
\affiliation{National Radio Astronomy Observatory, Charlottesville \\
VA 22903, USA}

\author[0000-0001-6196-4135]{Ralph~P.~Eatough}
\affiliation{National Astronomical Observatories, Chinese Academy of Sciences, 20A Datun Road, Chaoyang District\\
Beijing, 100101, China}
\affiliation{Max-Planck-Institut f{\"u}r Radioastronomie, 
Auf dem H{\"u}gel 69\\
D-53121 Bonn, Germany}

\author[0000-0002-9434-4773]{BoJun Wang}
\affiliation{National Astronomical Observatories, Chinese Academy of Sciences, 20A Datun Road, Chaoyang District\\
Beijing, 100101, China}

\author[0000-0003-1307-9435]{Paulo C. C. Freire}
\affiliation{Max-Planck-Institut f{\"u}r Radioastronomie, 
Auf dem H{\"u}gel 69\\
D-53121 Bonn, Germany}

\author{Kuo Liu}
\affiliation{Shanghai Astronomical Observatory, Chinese Academy of Sciences, NO. 80 Nandan Road\\
Shanghai, 200030, China}
\affiliation{Max-Planck-Institut f{\"u}r Radioastronomie, 
Auf dem H{\"u}gel 69\\
D-53121 Bonn, Germany}

\author{Zhen Yan}
\affiliation{Shanghai Astronomical Observatory, Chinese Academy of Sciences, NO. 80 Nandan Road\\
Shanghai, 200030, China}

\author{Jintao Luo}
\affiliation{National Time Service Center, Chinese Academy of Sciences \\ 
Xi'an, 710600, China}

\author{Liyun Zhang}
\affiliation{College of Physics, Guizhou University \\
Guiyang, 550025, China}

\author{Minghui Li}
\affiliation{State Key Laboratory of Public Big Data, Guizhou University \\
Guiyang, 550025, China}

\author[0000-0001-6051-3420]{Dejiang Yin}
\affiliation{College of Physics, Guizhou University \\
Guiyang, 550025, China}

\author{Baoda Li}
\affiliation{College of Physics, Guizhou University \\
Guiyang, 550025, China}

\author{Yifeng Li}
\affiliation{National Time Service Center, Chinese Academy of Sciences \\ 
Xi'an, 710600, China}

\author{Yinfeng Dai}
\affiliation{College of Physics, Guizhou University \\
Guiyang, 550025, China}

\author{Yaowei Li}
\affiliation{College of Physics, Guizhou University \\
Guiyang, 550025, China}

\author{Xinnan Zhang}
\affiliation{State Key Laboratory of Public Big Data, Guizhou University \\ 
Guiyang, 550025, China}

\author{Tong Liu}
\affiliation{National Astronomical Observatories, Chinese Academy of Sciences, 20A Datun Road, Chaoyang District\\
Beijing, 100101, China}

\author{Yu Pan$^{\dag}$}
\affiliation{School of Science, Chongqing University of Posts and Telecommunications \\
Chongqing, 40000, China}

\newcommand\blfootnote[1]{%
\begingroup 
\renewcommand\thefootnote{}\footnote{#1}%
\addtocounter{footnote}{-1}%
\endgroup 
}

\blfootnote{$\dag$  panyu@cqupt.edu.cn}


\begin{abstract}

We present the discovery of three pulsars in the Globular Cluster (GC) M15 (NGC 7078) by the Five-hundred-meter Aperture Spherical radio Telescope (FAST).
PSR~J2129+1210J (M15J) is a millisecond pulsar with a spin period of 11.84 ms and a dispersion measure of 66.68 pc cm$^{-3}$.
Both PSR~J2129+1210K and L (M15K and L) are long-period pulsars with spin periods of 1928 ms and 3961 ms, respectively.
M15L is the GC pulsar with the longest spin period known.
The timing solutions of M15A to M15H are updated. 
As predicted by \citet{ridolfi-2017}, 
the flux density of M15C keeps decreasing and the latest detection in our dataset was on December 20$^{\rm th}$, 2022.
We have also detected M15I's signal for the first time since its discovery. Current timing suggests that it is an isolated pulsar.

\end{abstract}



\section{Introduction} \label{sec:intro}
Pulsars are highly magnetized, rapidly-rotating neutron stars. The vast majority have been discovered with large single dish radio telescopes.
Since the discovery of the first Globular Cluster (GC) pulsar \citep[PSR B1821-24A,][]{lyne-1987}, 
GCs have proved to be prolific hosts of millisecond pulsars.
There are approximately 160 GCs in the Milky Way \citep{GC-pop}, 
and surveys for pulsars in these clusters have been performed for decades.
At the time of submission of this letter, 
a total of 330 pulsars have been identified in 44 GCs\footnote{Pulsars in globular clusters: \url{https://www3.mpifr-bonn.mpg.de/staff/pfreire/GCpsr.html}}.

\quad M15 (NGC 7078) is a core-collapsed GC \citep{harris-1996}.
It stands out as one of the oldest ($\sim 12.0\ {\rm Gyr}$) and most metal-poor Galactic GCs ($[{\rm Fe/H}] \sim -2.255$), 
with a significantly dense core \citep{sosin-1997,koleva-2008}. 
Observations of core-collapsed GCs like M15 often lead to surprising results, e.g. the discovery of an unusually slow GC pulsar with a period of 110.6 ms: M15A \citep{m15a}, and the first double neutron star system in a GC: M15C \citep{m15bc,deich-1996-M15Cmasses}. Until now, M15 had nine known pulsars associated with it,
with eight discovered in the 1980s and 1990s with the Arecibo telescope.
M15I was found by the Five-hundred-meter Aperture Spherical Radio Telescope \citep[FAST, ][]{FAST,2019SCPMA..6259502J} during a globular cluster pulsar survey\footnote{GC-FANS (Globular Cluster FAST: A Neutron-star Survey): \url{https://fast.bao.ac.cn/cms/article/65/}} \citep{pzc-24newpulsars}.

M15A is a bright and isolated pulsar, with a flux density of approximately $0.2 \pm 0.05$\,mJy at a frequency of 1415\,MHz. It was discovered in 1989 \citep{m15a}.
M15B was reported one year later, with a period of approximately 56.13 ms, exhibiting a flux density that is twice as faint as M15A, despite still being considered a bright isolated pulsar \citep{m15bc}.
M15C is in a highly eccentric binary system with an orbital period of 8 hours and an orbital eccentricity of 0.68 \citep{m15bc}.
Its companion is likely to be a neutron star, given its similarity to the well-known PSR~B1913+16 system \citep{weisberg_huang_2016}, but it has formed differently, in an exchange encounter that likely replaced the donor star that partially recycled the pulsar by the current companion \citep{prince_1991}. 

Several general relativistic effects have been measured in this system, which allowed precise mass measurements for its components and a test of general relativity~(GR): the observed orbital decay matches GR's prediction for gravitational wave damping of the orbit to within 1\% \citep{jacoby-2006}.
In previous studies, the brightness and polarimetric properties of M15C were reported, 
with a consistent decrease in flux density and significant changes in the (polarized) pulse profile \citep{kirsten-2014, ridolfi-2017}. 
This suggests that M15C's spin axis is precessing.
Therefore, regular observations of M15C remain important for understanding the effects of precession on the pulsar emission characteristics, 
the structure of M15C's emission beam, 
and the detection of potential signals originating from the companion of M15C.

After 2016, no follow-up observations of M15 have been reported.
The latest timing solutions for M15A, B, and C were reported in 2006 \citep{jacoby-2006}
while the ephemerides for the other pulsars have not been updated since the 1990s \citep{anderson-1993}.

There is a nearly thirty-year gap between the discovery of M15H and M15I. The reason for this is that the survey that found pulsars M15E-H was exceptionally deep, making use of the old Arecibo 430 MHz line feed, which had 
a uniquely large
gain of $19\,\rm{K\,Jy}^{-1}$ \citep{anderson-1993}. Most other Arecibo receivers, especially those in the Gregorian dome (which was built after the upgrade in the late 1990s) had gains of 
approximately $10\,\rm{K\,Jy}^{-1}$. The Gregorian receivers 
generally had uniform sensitivity at almost all zenith angles but they illuminated a smaller portion of the main reflector. While these receivers had less spillover from the ground, reducing the system temperature, for sources near the zenith, the old 430\,MHz line feed illuminated almost all the main reflector, and also did not pick up much of the ground spillover, making it the most sensitive reveiver at zenith.

The pulsars discovered in this survey were about the faintest among radio pulsars for decades; only now, with FAST, are we able to find even fainter pulsars.
M15I was discovered in 2021 \citep{pzc-24newpulsars} and has only been detected in one observation. 
It was still considered to be a new pulsar due to its strength and behavior in the observation: the signal-to-noise ratio (SNR) of its initial discovery is about 12.5 and its signal disappeared when FAST was not aligned with M15, and reappeared after realignment.
According to simulations, 
M15 is possibly one of the GCs with the highest number of pulsars \citep{bagchi-2011, turk-2013, yin_et_al+24}.
Therefore, a pulsar search within M15 is likely to find additional pulsars; for this reason,
since the beginning of the FAST GC pulsar surveys, M15 was one of the targets with the highest priority.
In this letter, we report the discoveries and timing solutions of three pulsars in M15, the re-detections of M15C and I, and updated timing solutions for M15A to H.

The structure of this letter is as follows: 
Section 2 presents the observations and data reduction, 
Section 3 provides the updated timing solutions,
the discussion is given in Section 4, 
and Section 5 is the summary.

\section{Observations and data reduction} \label{sec:obs}

With the aim of finding new pulsars, we started observing M15 with FAST in October 2018.
During the following six years, a total of 18 observations were performed, all with the L-band 19-beam receiver 
which covers a frequency range of 1 to 1.5 GHz with 4096 channels, 
corresponding to a channel width of 0.122 MHz.
All of the observational data were {\sc psrfits} files in pulsar search mode and only data from the central beam were used in this study.
The beamwidth is about 3$'$ at 1.4 GHz \citep{pzc-24newpulsars}.
The core radius and half-light radius of M15 are $0.14'$ and $1.00'$, respectively \citep{harris-1996}.
Thus, the central beam would be enough to cover the area where we would expect to detect signals from the majority of pulsars.


Most of the observations of M15 were made with the {\sc Tracking} mode of FAST where a single position is continuously tracked with the central beam of the 19-beam receiver for the full duration of the observation. One of the observations (on 2022.01.02) was conducted in the  {\sc Snapshot} mode where a grid of four adjacent pointings, each of 30\,minutes, is done to uniformly cover a wider region of diameter $\gtrsim 30'$. For this epoch only a single pointing from the {\sc Snapshot}  covering the cluster core was analysed. All of the FAST data up to January 31, 2023 has been released\footnote{\url{https://fast.bao.ac.cn/cms/article/275/}}.

To analyze the data, 
we employed the pulsar search code PRESTO \citep{presto-2011}.  
We dedispersed data by using {\sc prepsubband} in PRESTO.
The dispersion measure (DM) range of known pulsars in M15 is $\sim65.5$ to $67.7\,{\rm pc\,cm}^{-3}$.
Correspondingly, we dedispersed the data within a DM range of $65$ to $69\,{\rm pc\,cm}^{-3}$, with a DM step of $0.05\,{\rm pc\,cm^{-3}}$.
Subsequently, we applied {\sc realfft} to transform the dedispersed time series to the frequency domain and then applied {\sc accelsearch} for a periodic signal search.
Because the signal of binary M15C was one of our primary targets,
the acceleration search parameter {\sc zmax} was set to be 100.
{\sc zmax} is the maximum spectral drift in Fourier frequency bins, caused by a constant acceleration, that is searched.
The relation between the {\sc accelsearch} parameter {\sc zmax}, here denoted by $z$, and a pulsar's orbital acceleration, $a_{\rm l}$ is $z = \frac{T^{2}_{\rm obs}a_{l}f}{c}$,  where $T_{\rm obs}$ is the duration of the observation, $f$ is the pulsar's spin frequency and $c$ is the speed of light \citep{zmax}.
For example, for a pulsar spinning at $f\simeq200\,{\rm Hz}$ (like M15I) a spectral drift in the fundamental frequency of $z=100$ in a 1-hour observation would 
correspond to an acceleration, $a_{l}$, of $\sim 11.5\,{\rm m}\,{\rm s}^{-2}$ .
The number of harmonics summed was up to 32 and the search results were sifted using PRESTO's {\sc ACCEL\_}\text{sift.py}. According to the optimal DM and period values of the candidates found, the data were then folded with {\sc prepfold} from PRESTO.
To save time, we used RPPPS 4.0\footnote{\url{https://github.com/qianlivan/RPPPS}),\cite{RPPPS}} to run these PRESTO routines in parallel.

In a previous study, \citet{verbunt-2014-slowpulsar} discussed possible causes for the observed presence of slow pulsars in core-collapsed GCs, postulating that the large rate of stellar encounters per binary in such clusters causes the disruption of low-mass X-ray binaries, leading to the formation of partially recycled pulsars.
An early example was the discovery of M15A, and a recent, more extreme example is a slow pulsar with a $\rm \sim$2.4 s period found in the core-collapsed GC NGC~6624 \citep{2.4-pulsar}.
Such pulsars could be young neutron stars that have been produced by a different type of core-collapse mechanism\citep{ye-2024,kremer-2023}.
Being a core-collapsed GC, too, M15 may also contain pulsars that are significantly slower than M15A.
The Fast Folding Algorithm (FFA) is potentially a suitable technique to find new long period pulsars in M15.
The FFA code Riptide-FFA\footnote{\url{https://riptide-ffa.readthedocs.io/en/latest/}} was used in the data processing. 
In terms of FFA search settings, 
we followed the examples provided on the Riptide-FFA website\footnote{Riptide-FFA: \url{https://riptide-ffa.readthedocs.io/en/latest/pipeline.html\#number-of-parallel-processes}}.
The period ranges used for our search were divided into three stages, the first being for shorter periods of 0.1 to 1\,s, 
the second for intermediate periods of 1 to 5\,s, 
and the third for longer periods from 5 to 180\,s.
The search results from both PRESTO and Riptide-FFA are all presented in Table \ref{Table.1}. 

\quad
\begin{table}[H]
\centering
\setlength{\tabcolsep}{3pt}  
\caption{M15 pulsar detections. The “$\surd$” stands for those detections found in search results from either PRESTO, Riptide-FFA or both, “$\star$” are for those not detected in search results but can be found by folding with timing solutions. After the discoveries of M15J, M15K and M15L in the search results, we used either their barycentric periods or rough fits of the spin frequency (F0) to fold other observations. 
}

\begin{tabular}{lccccccccccccc}
  \hline
  \multicolumn{1}{l}{Obs. date} & \multicolumn{1}{l}{$T_{\rm obs}$\,(h)} & M15A & M15B & M15C & M15D & M15E & M15F & M15G & M15H & M15I & M15J & M15K & M15L \\
  \hline\hline
    2018.10.13 & 1.0  & $\surd$ & $\surd$ & $\surd$ & $\surd$ & $\surd$ & $\surd$ &         & $\star$ & $\star$ & $\star$ & $\star$ & $\star$ \\
    2018.11.17 & 1.1  & $\surd$ & $\surd$ & $\surd$ & $\surd$ & $\surd$ & $\star$ &         & $\star$ & $\star$ & $\star$ & $\star$ & $\star$ \\
    2019.04.25 & 0.7  & $\surd$ & $\surd$ & $\star$ & $\surd$ & $\surd$ & $\surd$ &         & $\surd$ &         &         &         &         \\
    2019.08.18 & 0.3  & $\surd$ & $\surd$ & $\star$ & $\surd$ & $\surd$ &               &         & $\star$ &          &         &         &         \\
    2019.11.09 & 1.3  & $\surd$ & $\surd$ & $\star$ & $\surd$ & $\surd$ & $\surd$ &         & $\surd$ &$\star$ & $\star$ & $\star$ & $\star$ \\
    2019.12.14 & 2.0  & $\surd$ & $\surd$ & $\star$ & $\surd$ & $\surd$ & $\surd$ &         & $\star$ &           & $\surd$ & $\star$ & $\star$ \\
    2020.08.30 & 0.7  & $\surd$ & $\surd$ & $\surd$ & $\surd$ & $\surd$ & $\surd$ &         & $\star$ &          &         &         & $\star$ \\
    2020.09.02 & 3.0  & $\surd$ & $\surd$ & $\star$ & $\surd$ & $\surd$ & $\surd$ &         & $\star$ & $\star$ & $\star$ &         & $\star$ \\
    2020.09.22 & 2.5  & $\surd$ & $\surd$ & $\star$ & $\surd$ & $\surd$ & $\surd$ &         & $\surd$ & $\surd$ &         & $\surd$ & $\surd$ \\
    2020.09.25 & 4.0  & $\surd$ & $\surd$ & $\star$ & $\surd$ & $\surd$ & $\surd$ &         & $\star$ & $\star$ & $\star$ &         & $\star$ \\
    2020.12.21 & 4.5  & $\surd$ & $\surd$ & $\star$ & $\surd$ & $\surd$ & $\surd$ &         & $\surd$ &$\star$&         & $\surd$ & $\surd$ \\
    2021.03.09 & 2.4  & $\surd$ & $\surd$ &         & $\surd$ & $\surd$ & $\surd$ & $\star$ & $\star$ &$\star$& $\star$ & $\star$ & $\star$ \\
    2022.01.02 & 0.5  & $\surd$ & $\surd$ &         & $\surd$ & $\surd$ & $\surd$ & $\surd$ & $\star$ & $\star$&         & $\star$ &         \\
    2022.07.07 & 0.9  & $\surd$ & $\surd$ &         & $\surd$ & $\surd$ & $\surd$ & $\star$ & $\star$ & $\star$&         & $\star$ & $\star$ \\
    2022.11.19 & 1.8  & $\surd$ & $\surd$ &         & $\surd$ & $\surd$ & $\surd$ & $\star$ & $\surd$ & $\star$& $\star$ & $\star$ & $\star$ \\
    2022.12.20 & 2.8  & $\surd$ & $\surd$ & $\star$ & $\surd$ & $\surd$ & $\surd$ & $\star$ & $\surd$ &$\star$ & $\star$ & $\star$ & $\star$ \\
    2023.01.20 & 2.8  & $\surd$ & $\surd$ &         & $\surd$ & $\surd$ & $\surd$ & $\star$ & $\surd$ & $\surd$ & $\star$ & $\star$ & $\star$ \\
    2023.02.20 & 0.7  & $\surd$ & $\surd$ &         & $\surd$ & $\surd$ & $\surd$ &         & $\star$          &           & $\star$ & $\star$ & \\
  \hline
  Total &  & 18 & 18 & 12 & 18 & 18 & 17 & 6 & 18 & 13 & 11 & 13 & 14 \\
  \hline
\end{tabular}
\label{Table.1}
\end{table}

\section{Results} \label{sec:Res}

The 18 observations were carried out from October 2018 (MJD 58404) to February 2023 (MJD 59995), 
spanning 1591 days.
Due to variations in observation scheduling, 
the observations lasted for half an hour to five hours. 

\subsection{New Pulsars}

\subsubsection{M15J}
From the results obtained with Riptide-FFA, a signal with a SNR of 9.2 was detected in the data recorded on December 14, 2019.
Its pulse profile had 10 peaks with a period of 118.43\,ms and an optimal
DM of 66.40 $\text{pc\,cm}^{-3}$.
As indicated by the presence of multiple profile peaks, subsequent investigations revealed that the fundamental period of this signal was actually one tenth of this, at $11.84\,{\rm ms}$. This is the newly discovered pulsar, M15J and our original detection with Riptide-FFA was through its tenth sub-harmonic.



The pulsar was successfully detected in multiple subsequent observations.
With more data recently obtained by FAST, 
the phase-connected timing solution was obtained (see Table.~\ref{Table.2}). 
M15J is thus an isolated pulsar with a spin period of 11.84 ms and with a DM value of 66.68 pc cm$^{-3}$.
\subsubsection{M15K \& M15L}
From the observation of December 21$^{\rm st}$, 2020, two long-period pulsars, namely M15K and M15L, were detected by Riptide-FFA. 
The signals of M15K and M15L have periods of approximately 1928 ms and 3961 ms,
with trial DM values of 66.50 and 66.75 $\text{pc\,cm}^{-3}$ for the best SNR, respectively.
Due to Radio Frequency Interference (RFI), their SNRs are mostly very low.
In order to detect them more often, more weak RFI should be excised from the data. Timing solutions for these two pulsars are presented in Table~\ref{Table.2}.


Recently another independent study also reported the discoveries of these two pulsars with their phase-connected
timing solutions. Their results show that both pulsars are below the spin-up line, indicating that the pulsars could be partially recycled \citep{Zhou-2024}.

\subsubsection{M15M \& M15N}
In addition, another two new pulsars named M15M and M15N have been discovered by stacking power spectra together. 
They are likely to be isolated MSPs with periods of about 4.83\,ms and 9.28\,ms respectively. More discovery details will be presented in upcoming papers (Dai~et~al.,~in prep).

\subsection{Known Pulsars}
M15C appeared only three times in the search results.
To obtain as many detections as possible, we also used the ephemeris from the ATNF Pulsar Catalogue\footnote{psrcat, \url{https://www.atnf.csiro.au/research/pulsar/psrcat/}} to fold (with PRESTO {\sc prepfold}) the individual observations \citep{ATNF}. Using these folding results we constructed a new timing solution.
%
The fitted orbital period, projected semi-major axis, epoch of periastron, eccentricity, and longitude of periastron are 
0.3352820162(3) days, 2.51842(1) lt-s, 50000.064743(8),  0.681404(9) and 345.312(2) deg, 
being consistent with previous results \citep[e.g.][]{anderson-1993}.

We have detected all the other known pulsars in the GC; their
detection rates can be found in Table \ref{Table.1}.
We have obtained the fully phase-connected timing solutions for M15A, B, D, E, F, G and H (see Table~\ref{Table.2}, ephemerides from the ATNF Pulsar Catalogue were also used for the timing of M15G and H).
For M15G, its profile appeared notably complex upon its discovery \citep{anderson-1993}.
Using {\sc dracula}\footnote{\url{https://github.com/pfreire163/Dracula}}\citep{dracula}, we were able to derive a unique timing solution for this pulsar that is basically consistent with that presented by \citet{anderson-1993}. Thus, 
we believe the signals detected are from M15G.
The timing solutions for M15A, B, D, E, F, G and H are basically consistent with previous studies. 

\begin{figure}[H]
  \centering
  \includegraphics[width=0.99\textwidth]{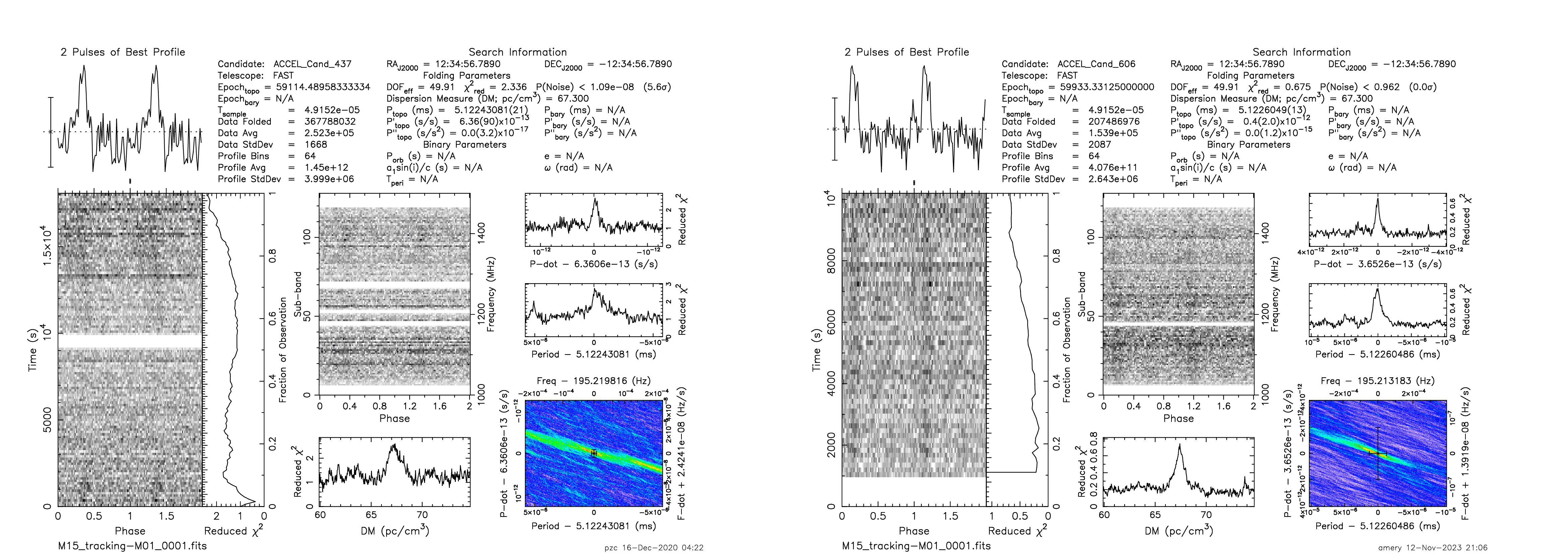}
  \caption{M15I detections, on MJD 59144 (left) and MJD 59933 (right).  These pulsar candidates heuristic plots were generated with the PRESO {\sc prepfold} routine. Starting clockwise from the bottom left, the first panel shows the pulse profile as a function of observing time through 128 subintegrations represented in grayscale, as well as the reduced $\chi^2$ -- a measure of the profile significance -- also as a function of time. White banding are subintegrations excised due to radio frequency intereference (RFI). On the top left is the integrated pulse profile. To the right are the various discovery parameters in the topocentric and barycentric reference frames. The three panels on the right show the results of the period and period derivative optimisation process. The lower middle panel shows the reduced $\chi^2$  as a function of trial dispersion measure, and the grayscale panel in the middle shows the pulse profile as a function of observing frequency across 128 subbands where subbands excised due to RFI are visible as white bands.} 
  \label{Fig1}
\end{figure}

M15I was detected one more time on December 20\textsuperscript{th}, 2022 (MJD 59933).
In comparison to its discovery and confirmation observation on September 22$^{\rm nd}$ 2020 (MJD~59114),
the signal of M15I on December 20$^{\rm th}$ 2022 matched well both in the spin period and the shape of the pulse profile (see Figure~\ref{Fig1}).
From the observation on September~22$^{\rm{nd}}$~2020, 
its barycentric period and the acceleration inferred from the period derivative are 5.1221966(4) ms and -0.011(11) ${\rm m\ s}^{-2}$, respectively, 
while they are 5.1221974(2) ms and 0.0071(36) ${\rm m\ s}^{-2}$, respectively, from the observation on December 20$^{\rm th}$ 2022. Despite such a small but notable difference, our current timing results suggests it is an isolated pulsar. But due to insufficient detections, we did not obtain its phase-connected timing solution. Therefore, we only fitted F0 for M15I (see Table~\ref{Table.2}). Further details will be presented in Section~4.4.


\begin{table}[]
\caption{Measured parameters for the pulsars in GC M15 from FAST. Timing position for M15I was not obtained and its position is listed as the center of the GC marked with *. The solar system ephemeris is DE421 for all the timing solutions.}
\label{Table.2}
\centering
\scriptsize
\setlength{\tabcolsep}{6pt}
\renewcommand{\arraystretch}{1.4}
\begin{tabular}{l c c c c}
\hline
Pulsar & 2129+1210A & 2129+1210B & 2129+1210C & 2129+1210D \\
\hline\hline
Right Ascension, $\alpha$ (J2000) & 21:29:58.24642(4) &  21:29:58.6280(1) & 21:30:01.2031(6) & 21:29:58.2692(1) \\
Declination, $\delta$ (J2000) & 12:10:01.172(1) & 12:10:00.283(7) & 12:10:38.19(2) & 12:09:59.645(2) \\
spin Frequency, $f$ (s$^{-1}$) & 9.0363060726718(5) & 17.814819156216(5) & 32.7554227003(4) & 208.21173107090(3) \\
1st spin Frequency derivative, $\dot{f}$ (Hz s$^{-2}$) & $1.713989(9) \times 10^{-15}$ & $-3.02952(6) \times 10^{-15}$ & $-5.3523(5) \times 10^{-15}$ & $4.6391(4) \times 10^{-14}$ \\
Reference Epoch (MJD) & 58598.012893 & 58404.479166 & 58404.483246 & 58831.3534 \\
Start of Timing Data (MJD) & 58404.479 & 58404.479 & 58404.479 & 58404.479 \\
End of Timing Data (MJD) & 59247.313 & 59964.318 & 59964.318 & 59964.329 \\
Dispersion Measure, DM (pc cm$^{-3}$) & 67.226(1) & 67.731(6) & 67.10 & 67.277(2) \\
Number of TOAs & 219 & 252 & 109 & 248 \\
Weighted rms timing residual ($\mu$s) & 27 & 72 & 99 & 38 \\
\hline

PULSAR & 2129+1210E & 2129+1210F & 2129+1210G & 2129+1210H \\
\hline\hline
Right Ascension, $\alpha$ (J2000) &21:29:58.18447(4)& 21:29:57.1783(1) & 21:29:57.964(8) & 21:29:58.1828(3)  \\
Declination, $\delta$ (J2000) &12:10:08.508(1)& 12:10:02.818(2) & 12:09:56.5(2)& 12:09:59.291(5)  \\
spin Frequency, $f$ (s$^{-1}$) &214.987399714593(8)& 248.32119058101(2) & 26.553254549(6) &148.2932725188(7)  \\
1st spin Frequency derivative, $\dot{f}$ (Hz s$^{-2}$) &$-8.5721(1) \times 10^{-15}$& $-1.6719(4) \times 10^{-15}$ & $-1.152(6) \times 10^{-15}$ & $-4.702(7) \times 10^{-16}$  \\
Reference Epoch (MJD) & 59883.353 & 59933.331 & 47632.520& 47632.520 \\
Start of Timing Data (MJD) & 58404.483 & 59404.483 & 59282.167&58796.482 \\
End of Timing Data (MJD) & 59995.172 & 59995.172& 60357.237& 59964.324 \\
Dispersion Measure, DM (pc cm$^{-3}$) &66.586(1)& 65.597(2) &66.40 & 67.117(5)  \\
Number of TOAs & 247 & 165 & 29 &167  \\
Weighted rms timing residual ($\mu$s) & 18 & 22 & 598 & 94  \\
\hline

PULSAR & 2129+1210I & 2129+1210J & 2129+1210K & 2129+1210L \\
\hline\hline
Right Ascension, $\alpha$ (J2000) & 21:29:58.33$^{*}$ & 21:29:58.4127(4) & 21:29:58.43(4) & 21:29:58.2(1) \\
Declination, $\delta$ (J2000) & 12:10:01.2$^{*}$ & 12:10:16.014(9) & 12:10:00(1) & 12:09:33(2) \\
spin Frequency, $f$ (s$^{-1}$) & 195.228710(2) & 84.44174596102(2) & 0.51855097775(1) & 0.25247957713(1) \\
1st spin Frequency derivative, $\dot{f}$ (Hz s$^{-2}$) & ...... & $-1.4551(5) \times 10^{-15}$ & $-3.174(2) \times 10^{-16}$ & $-5.57(2) \times 10^{-15}$ \\
Reference Epoch (MJD) & 59995.156 & 59964.210 & 60000 & 60000 \\
Start of Timing Data (MJD) & 58094.647 & 58796.482 & 58404.510 & 58404.510 \\
End of Timing Data (MJD) & 59995.191 & 60408.110 & 60408.010 & 60408.010 \\
Dispersion Measure, DM (pc cm$^{-3}$) & 67.50 & 66.688(8) & 66.50 & 66.10 \\
Number of TOAs & 57 & 210 & 73 & 54 \\
Weighted rms timing residual ($\mu$s) & 88 & 60 & 6300 & 12000 \\
\hline
\end{tabular}
\end{table}

\section{Discussion} \label{sec:dis}

\begin{figure}[]
  \begin{center}
  \includegraphics[width=0.9\textwidth]{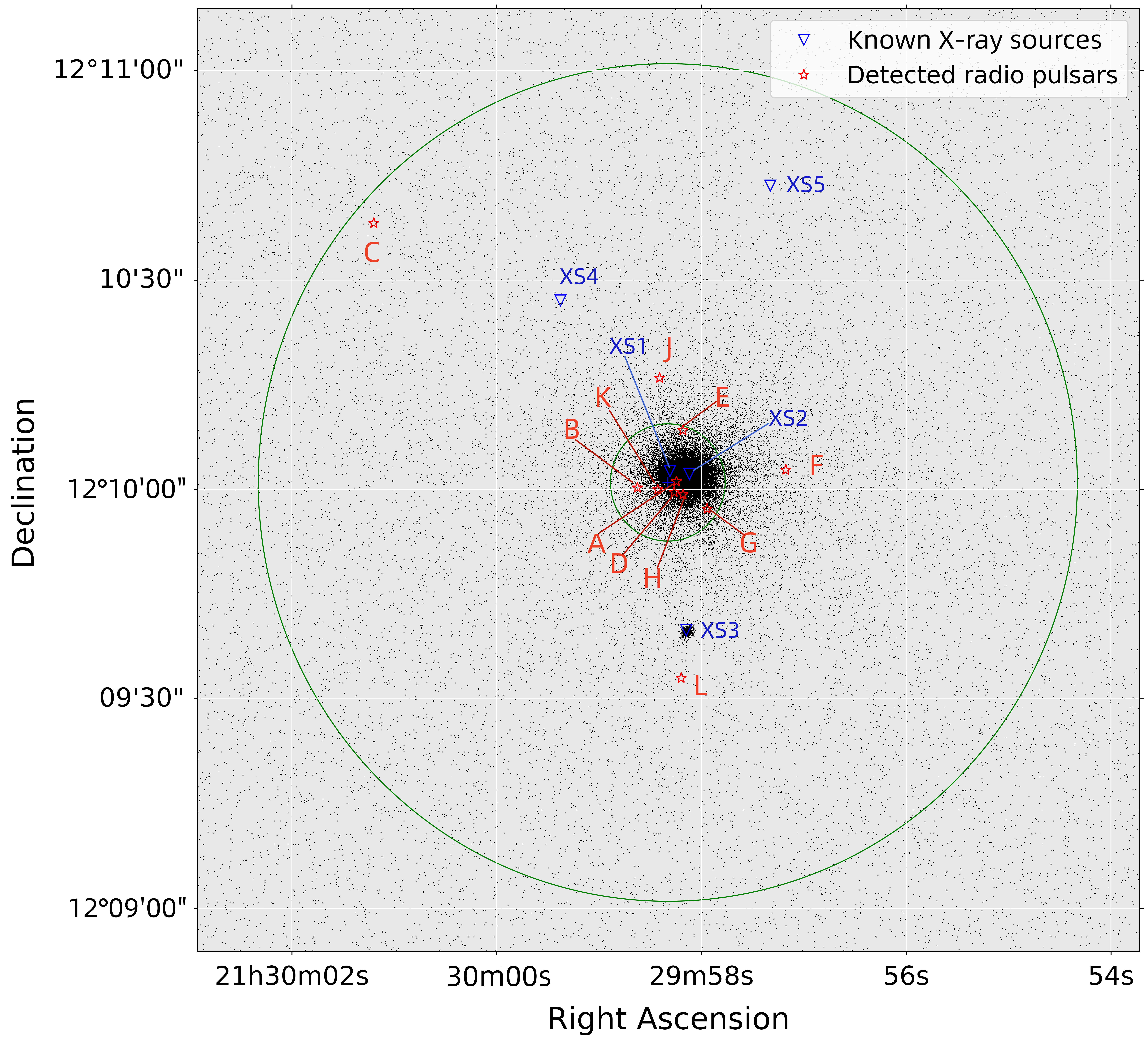} 
  \caption{X-ray image of M15 in the energy range 0.2 to 20\,keV from {\it Chandra} archival data. The blue triangles show the positions of known X-ray sources while the red stars indicate the positions of the known radio pulsars. The large green circle and smaller green circle indicate the half-light radius and core radius of M15 respectively \citep{harris-1996}.\label{Fig2}}
  \end{center}
\end{figure}

In Figure~\ref{Fig2}, the pulsar timing positions and {\it Chandra} X-ray source positions in M15 are presented on an X-ray map from the {\it Chandra} Data Archive (Obs ID: 14618) in equatorial coordinates.
Among these pulsars, none of them has an X-ray counterpart.
MSPs can emit X-ray radiation in different ways: thermal origin, non-thermal origin and emission from a shock as seen in ``black widow" and ``redback" pulsars
\citep{harding-thermal,berker-xray,blackwidowxray,redbackxray}.
Unlike what is often seen in such {\it spider} systems none of the pulsars in M15 show eclipses.
The non-detection of the X-ray counterpart indicates the thermal radiation from M15 pulsars is too faint to be detected, 
or the axis of non-thermal radiation may not be aligned with the Earth.

\begin{figure}[]
  \centering
  \includegraphics[width=0.95\textwidth]{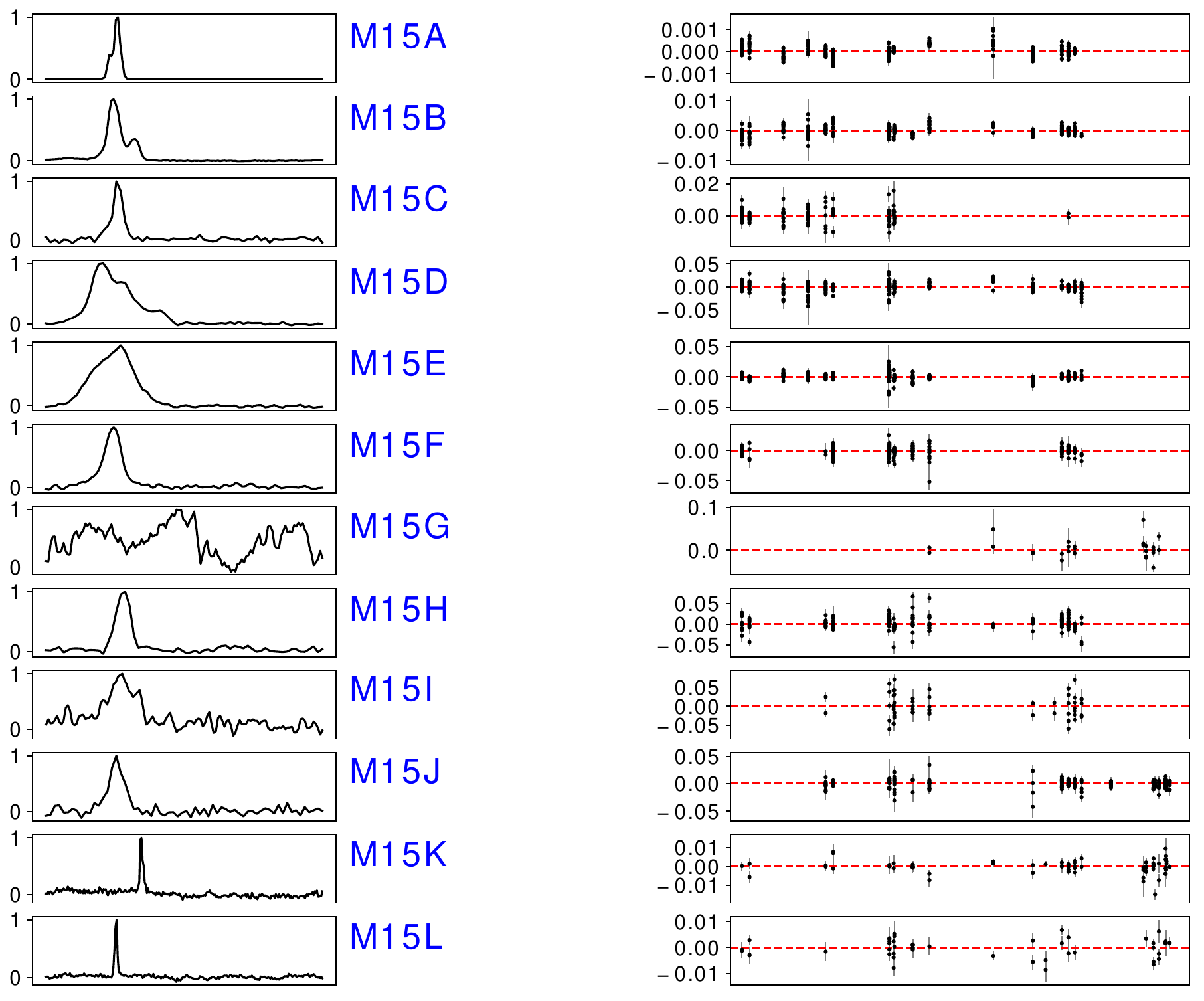} 
  \caption{Average pulse profiles({\it left}) and timing residuals for the pulsars in M15({\it right}, the MJD of the timing residuals spans from 58350 to 60500). In all profiles the uncalibrated peak flux density is normalized to unity. The Y axis of the residual plot is 
  in units of pulse phase.
  To obtain better timing solutions for the new pulsars and M15G, we used TOAs from the most recent observation data. For the other pulsars, their timing solutions are already well-established, so we did not include the newest data (observations after MJD 60000) for them.
  The profile of M15G appeared notably complex upon its discovery. In our study, we were able to derive a timing solution for M15G according to these detections.}
  \label{Fig3}
\end{figure}

In Figure \ref{Fig3}, the integrated average pulse profiles and timing residuals for all the pulsars in M15 are presented.
Due to limitations in data quality, not all observational data were utilized.

\subsection{Detection Rates}
Due to interstellar scintillation, the flux densities of PSRs M15A, B, D, and E changed between observations. Because the calibration noise diode was used infrequently, our current dataset does not allow us to accurately calibrate their flux densities at all epochs. Instead we used the radiometer equation to estimate their flux:
\begin{equation}
    S_{\rm mean}=\frac{({\rm SNR})\beta T_{\rm sys}}{G\sqrt[]{n_{\rm p}t_{\rm obs}\Delta f}} \sqrt[]{\frac{W}{P-W}}
\end{equation} 
among which $\beta$ is the sampling efficiency and the value of it is 1 for our 8-bit recording system; the system temperature ($T_{\rm sys}$) is $24{\rm K}$; the antenna gain ($G$) is $16\,{\rm K\,Jy}^{-1}$; the number of polarizations ($n_{\rm p}$) is 2; the $t_{\rm obs}$ is the integration time in units of seconds; the $\Delta f$ is the bandwidth, here is 300\,MHz; $P$ is the pulsar period and $W$ is the pulse width.

From the estimation, the rough flux density range of our detections is 0.094 to 0.21~mJy for M15A, 0.011 to 0.052~mJy for M15B, 0.017 to 0.056~mJy for M15D and 0.012 to 0.033~mJy for M15E. They are all 
higher than the predicted FAST sensitivity limit in M15 observations of about 
0.0005\,mJy (this assumes that the pulsar's duty cycle was 10\% and the integration time was 5 hours)~\citep{pzc-24newpulsars}.

These four pulsars are bright enough to be detected even from the half-hour datasets.
Their timing solutions are consistent with those from previous studies \citep{anderson-1993}.
The timing result of M15C will be discussed in the next subsection. 

M15G, which was redetected only once in FAST data on January 02, 2022 \citep{pzc-24newpulsars}, now has 11 detections in total.
M15H was detected 7 times among the 18 observations (39\%) in the search results.
The low detection rate for M15H is due to the variation of its flux density likely caused by interstellar scintillation, but for M15H and other MSPs in M15, it was not showing a gradually weakening trend like M15C.
When folded with the timing solution, its signal appears in every observation.

M15I was re-detected the first time since its discovery, 
while M15J, an MSP, was detected 11 times (a detection rate of 61\%).

The pulse profile\textcolor{red}{s} of M15K and M15L were highly dependent on the RFI removal.
Nevertheless these observations were used to extract TOAs which were used to make timing solutions.

\subsection{The Double Neutron Star M15C}
M15C is in a well-studied pulsar-neutron star system. 
The mass of its companion and orbital parameters have all been obtained in previous studies \citep{anderson-1993,deich-1996-M15Cmasses,jacoby-2006}.
Compared with the results from 2014, M15C displayed much higher linear polarization in 2016 \citep{ridolfi-2017}.
It is predicted that the flux density of M15C will become fainter and finally disappear between 2041 and 2053.
In our dataset, the second most recent detection of M15C was on December 20$^{\rm th}$, 2022.
The duration of this observation is about three hours. 
The SNR of the M15C signal remained low (about 6.9). 
After this observation, 
no signals from M15C were detected 
until our three hour observation on January~20$^{\rm th}$, 2023.

\begin{figure}[]
  \centering
  \includegraphics[width=0.95\textwidth]{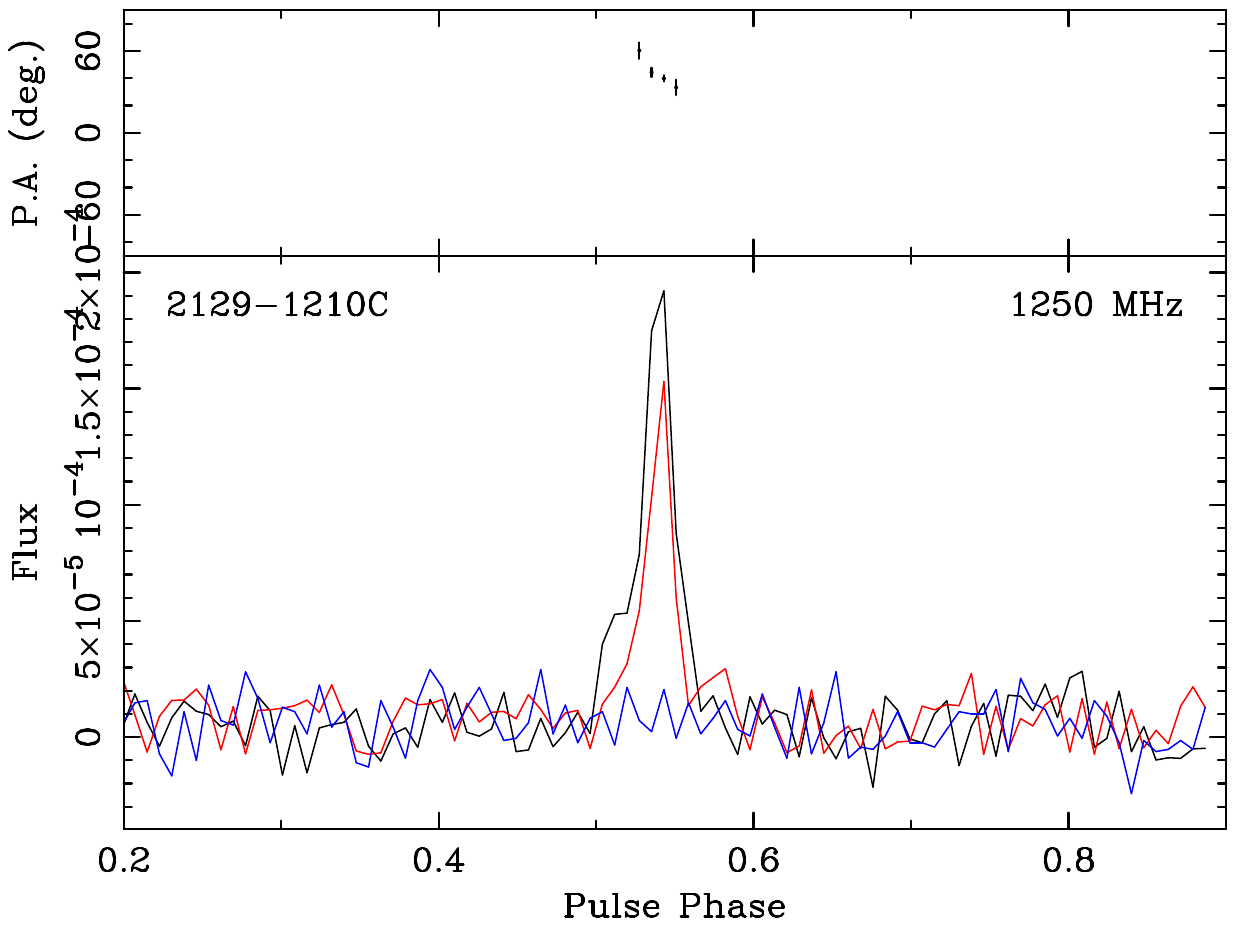}
  \caption{The lower panel shows the polarization calibrated pulse profile of PSR~M15C measured at FAST in 2019 - MJD~58829 (the black, red and blue lines represent total intensity, linear polarization and circular polarization respectively).  Measurements of the polarization position angle are shown in the upper panel.}
  \label{Fig4}
\end{figure}

A large part of our dataset were conducted only for pulsar searching, 
which consists of two summed polarizations and no calibration signal 
injection (from a noise diode). 
In order to perform polarization analysis, 
the data must be recorded with four polarization products and with the noise diode on at the beginning and end of the observation.
We only have three such observations.
Luckily, in the data obtained on December 14$^{\rm th}$, 2019, a signal from M15C was strong enough to perform a polarization analysis.
It is consistent with the previous study from \citet{ridolfi-2017} that shows M15C continues to exhibit a high degree of linear polarization (see Figure~\ref{Fig4}).



At L-band, the flux of pulsars with DM $\sim$ 68 pc cm$^{-3}$ changes every time we observe them due to interstellar scintillation \citep{gitika}.
In our study, interstellar scintillation does contribute to the variations of M15C's signal strength.
Due to limitations in our dataset, we are unable to precisely quantify the flux desnsity of pulsars in M15.
Therefore, we also used the radiometer equation to estimate the flux density M15C (Equation~1).


\setlength{\intextsep}{15pt}
\begin{table}[H]
\centering
\caption{Estimated flux density of M15C.}
\begin{tabular}{|l|c|c|}
  \hline
  \textbf{Obs. date} & \textbf{$T_{\rm obs}$\,(h)} & \textbf{Flux Density (mJy)} \\
  \hline\hline
  2018.10.13 & 1.0 & 0.0025 \\
  2018.11.17 & 1.1 & 0.0067 \\
  2019.04.25 & 0.7 & 0.0030 \\
  2019.08.18 & 0.3 & 0.0057 \\
  2019.11.09 & 1.3 & 0.0025 \\
  2019.12.14 & 2.0 & 0.0034 \\
  2020.08.30 & 0.7 & 0.0027 \\
  2020.09.02 & 3.0 & 0.0016 \\
  2020.09.22 & 2.5 & 0.0026 \\
  2020.09.25 & 4.0 & 0.0026 \\
  2020.12.21 & 4.5 & 0.0022 \\
  2021.03.09 & 2.4 & NO DETECTION \\
  2022.01.02 & 0.5 & NO DETECTION \\
  2022.07.07 & 0.9 & NO DETECTION \\
  2022.11.19 & 1.8 & NO DETECTION \\
  2022.12.20 & 2.8 & 0.0015 \\
  2023.01.20 & 2.8 & NO DETECTION \\
  2023.02.20 & 0.7 & NO DETECTION \\
  \hline
\end{tabular}
\label{tab:3}
\end{table}
At L-band, the signal of M15C is becoming increasingly faint and almost can not be detected after the conclusion of 2022.
The precise flux density measurements for pulsars in M15 are part of our ongoing GC pulsar flux measurement program, which will be discussed in a subsequent paper (Liu~et~al.,~in~prep).

\subsection{Slow Pulsars M15K and M15L}
In previous surveys, there were two pulsars in GCs that have been identified with periods more than one second: PSR~B1718--19A  \citep[1004 ms, NGC~6342A,][]{B17-18} and PSR~J1823--3022 \citep[2497 ms, 
thought to be in NGC 6624,][]{2.4-pulsar}.
PSR~B1718--19A is in a binary system with a low-mass non-degenerate star as its companion, 
associated with the core-collapsed GC, NGC~6342; while J1823--3022, due to the DM difference ($\sim$ 9.3 pc cm$^{-3}$ from the average DM for known pulsars) and an offset of 3 arcminutes from the GC center, could be unassociated with the core-collapsed GC NGC 6624.


The spin periods of M15K and L are 1.9 s and 3.9 s, respectively, 
ranking 3$^{\rm rd}$ and 1$^{\rm st}$ longest among all the GC pulsars.
In the other independent study which also reported on the discoveries of these two slow pulsars, their phase-connected timing solutions indicate both are below the spin-up line \citep{Zhou-2024}.
We also successfully obtained the phase-connected timing solutions for M15K and M15L. We followed the DM value by \citet{Zhou-2024} and re-fitted the other timing parameters with our dataset. The result of M15K has smaller errors, but the timing parameters of M15L show a tiny shift in position 
(e.g. the R.A. is 21:29:58.2(1) from this work and 21:29:57.92(9) in \citet{Zhou-2024}). Based on the positions and DM values, both M15K and L can be confirmed to be associated with GC M15.

M15K and L's profile baselines, and hence detections and TOAs are severely affected by significant RFI.
Eliminating RFI to enhance the signal quality and optimize their timing solution is the primary objective of our future work.


\subsection{Confirmation of M15I}


M15I is a millisecond pulsar that was detected and confirmed within a single observation.
Typically, confirmation of a pulsar requires detection in multiple observations.
However, during the observation of M15 on September 22, 2020, the FAST telescope encountered a mechanical issue, causing it to lose alignment with M15 for approximately 7000 seconds.
The signal from M15I was detected in this observation but disappeared during the 7000-second interval when the telescope was misaligned. Furthermore, when the telescope's alignment was corrected, the signal reappeared.
This indicated that this signal was indeed originated from GC M15.
Consequently, M15I was both detected and confirmed within a single observation, deviating from the conventional method.
A similar technique, known as spatial modulation search, was proposed for detecting and confirming highly nulling pulsars \citep{Qian-confirm}.
Our detection of M15I establishes the existence of M15I and validates the efficacy of this method.
With {\sc accelsearch} from PRESTO, M15I was detected again with the {\sc zmax} value of 100 
and with a DM value of 67.6 pc cm$^{-3}$.
M15I was detected at zero acceleration.
M15I was believed to be a binary\citep{pzc-24newpulsars}, however, we found that we can gain an adequate timing solution as an isolated pulsar.

\section{Conclusions}

In this paper, we presented our studies on the pulsars of the core-collapsed GC M15. 
The conclusions are as follows:

1, Observations to M15 had been made with FAST since 2018, 
resulting in re-detection for all the 9 previously known pulsars and the discovery of M15J, K, and L.

2, Updated timing solutions for known pulsars are consistent with previous ones \citep{anderson-1993} 

3, M15C, the double neutron star binary system, was detectable by FAST until December 20$^{\rm th}$, 2022. 
Its flux density kept decreasing. 
This may be the result of precession of the beam out of our line of sight.
The high degree of linear polarisation observed in late
2019 is consistent with that seen in earlier epochs. Its pulse profile is still highly linear polarized in the observation taken in December 14$^{\rm th}$, 2019.
As it disappears in FAST data, 
it may be possible to receive the signal from the companion and confirm if it is also a pulsar.

4, M15I was re-detected the first time since its discovery, current result indicates it could be an isolated pulsar.

5, Newly discovered pulsar M15J is an isolated pulsar, 
with a spin period of 11.84 ms and DM value of 66.68 pc cm$^{-3}$.

6, The two long period pulsars, M15K and L, have spin periods of 1.98 s and 3.61 s.
M15L is the GC pulsar with the longest spin period now known. 
The positions from the timing solutions suggest they are both associated with GC M15.

\begin{acknowledgments}
We sincerely appreciate the referee's  very careful review and the many constructive suggestions, which have significantly improved the quality of this work.
We also sincerely appreciate the editor’s continued support throughout this process.
This work is supported by the Natural Science Foundation of Chongqing No. cstc2021jcyj-msxmX0481.
Zhichen Pan and Lei Qian are supported by the CAS “Light of West China” Program.
Zhichen Pan and Lei Qian are also supported by the Youth Innovation Promotion Association of the Chinese Academy of Sciences (ID nos. 2023064 and 2018075,  Y2022027). RPE is supported by the Chinese Academy of Sciences President's International Fellowship Initiative, Grant No. 2021FSM0004. The National Radio Astronomy Observatory is a facility of the National Science Foundation operated under cooperative agreement by Associated Universities, Inc. SMR is a CIFAR Fellow and is supported by the NSF Physics Frontiers Center award 2020265.
\end{acknowledgments}

\bibliography{sample631}

\end{document}